\documentclass[aps,pre,reprint,superscriptaddress]{revtex4-1}
\usepackage{graphicx}
\usepackage{dcolumn}
\usepackage{bm}
\usepackage{graphicx}
\usepackage{dcolumn}
\usepackage{bm}
\usepackage{subfigure}
\usepackage{color}
\usepackage{amsmath}

\begin{document}


\renewcommand{\thetable}{\arabic{table}}

\newcommand{\change}[1]{\textcolor{red}{#1}} 
\newcommand{\BU}{Department of Mechanical Engineering, Division of Materials Science and Engineering, and the Photonics Center, Boston University, Boston, Massachusetts 02215, USA}
\newcommand{\Harvard}{Harvard John A. Paulson School of Engineering and Applied Sciences,\\
 Harvard University, Cambridge, MA 02138, USA}


\title{Nanofluidics of Single-crystal Diamond Nanomechanical Resonators}

\author{V. Kara}
\affiliation{\BU}

\author{Y.-I. Sohn}
\affiliation{\Harvard}

\author{H. Atikian}
\affiliation{\Harvard}

\author{V. Yakhot}
\affiliation{\BU}

\author{M. Lon\u{c}ar}
\affiliation{\Harvard}

\author{K. L. Ekinci}
\email[Electronic mail: ]{ekinci@bu.edu}
\affiliation{\BU}

\date{\today}

\begin{abstract}

Single-crystal diamond nanomechanical resonators are being developed for countless applications. A number of these applications require that the resonator be operated in a fluid, i.e., a gas or a liquid. Here, we investigate the fluid dynamics of single-crystal  diamond nanomechanical resonators in the form of nanocantilevers. First, we measure the pressure-dependent dissipation of  diamond  nanocantilevers with different linear dimensions and frequencies in three gases, He, N$_2$, and Ar. We observe that a subtle interplay between the length scale and the frequency  governs the scaling of the fluidic dissipation. Second, we obtain a comparison of the surface accommodation of different gases on the diamond surface by analyzing the dissipation in the molecular flow regime. Finally, we measure the thermal fluctuations of the nanocantilevers in water, and compare the observed dissipation and frequency shifts with theoretical predictions. These findings set the stage for developing diamond nanomechanical resonators operable in fluids.


\end{abstract}

\maketitle

\section{Introduction}

Single-crystal diamond has unique and attractive mechanical properties, such as a high Young's modulus, a high thermal conductivity, and a low intrinsic dissipation. Recent advances in  growth and nanofabrication techniques have allowed for the  fabrication and operation of nanometer scale mechanical systems made out of  single-crystal diamond. Part of the research community in diamond nanomechanics is focused on coupling the negatively charged nitrogen vacancy (NV$^-$) with a mechanical degree of freedom \cite{Ovartchaiyapong, Teissier, MacQuarrie_2,Barfuss}.  There are significant efforts for realizing diamond nano-opto-mechanical systems \cite{Khanaliloo,Burek,Sohn} for quantum information processing and optomechanics. Diamond nanocantilevers are also being developed for ultrasensitive magnetometry \cite{Maletinsky,Grinolds} and scanning probe microscopy (SPM) \cite{K_Kim}. In addition, it has  been suggested that the  chemistry of the diamond surface may be amenable to surface functionalization \cite{Wensha_Yang} for sensing.

\begin{figure}
\centering
\includegraphics[width=3.375in]{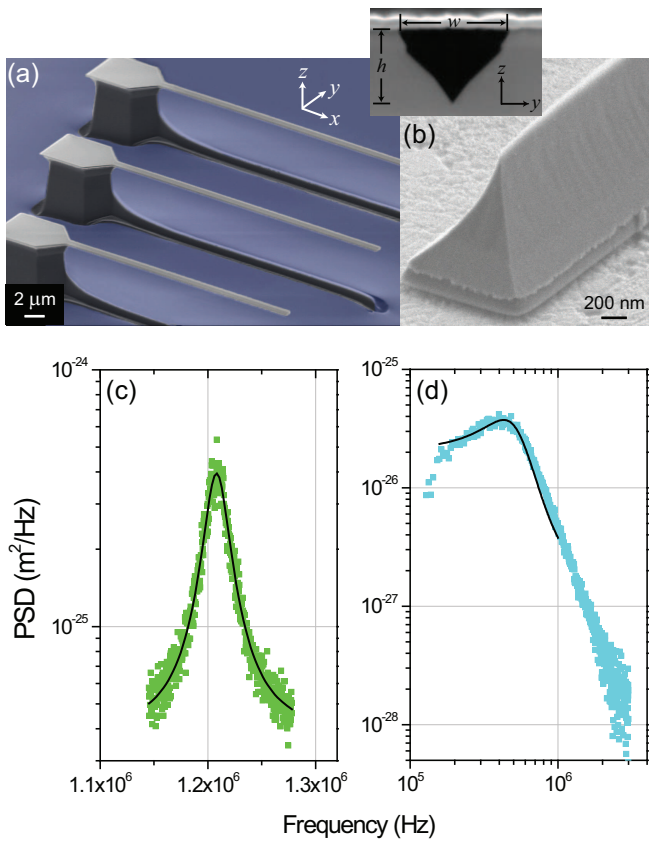}
\caption{(a) SEM image of a set of nanomechanical cantilever resonators with the same cross-sectional dimensions but different lengths. The gap from the bottom of each nanocantilever to the substrate is $\approx 6~\mu\rm m$.  (b) Image of the sidewalls of a nanocantilever. The inset shows a tilt-corrected  image of the cross-section with  dimensions $w\times h \approx \rm 820~ nm \times 530 ~nm$. (c)-(d) Power spectral density (PSD) of the thermal fluctuations of a nanocantilever  ($l \times w \times h = 29 \times 0.820 \times 0.530~\mu {\rm m}^3$, and vacuum  parameters $f_0 \approx 1.211$ MHz and $Q_0\approx 10^5$) in fluids. The solid lines are fits to Lorentzians. (c) In atmosphere, the frequency and quality factor become $f_{atm}\approx 1.208$ MHz and $Q_{atm} \approx 10^2$, respectively. (d) In water, $f_w \approx 0.43$ MHz and $Q_w\approx 1.1$ }
\label{fig:experimental}
\end{figure}

To date, the performance of single-crystal diamond nanomechanical resonators  has been thoroughly evaluated in vacuum \cite{Tao, Sohn, Ovartchaiyapong, Burek}. However,  vacuum properties will  be of  little relevance for some applications, such as mass sensing, magnetometry and SPM. Instead,  performance in fluids will be consequential --- especially, when biological or chemical samples are being analyzed in ambient air or in liquids. It is therefore important to elucidate the nanoscale fluid dynamics (or nanofluidics) of diamond nanomechanical resonators.  The smooth and inert surface of single-crystal diamond may provide unique opportunities for high-performance operation in fluids. For instance, gases may be accommodated favorably on the diamond surface; the inherent  inertness of the diamond surface may allow for reduced drag in water \cite{Sukumar_PRL}.

In this manuscript, we present a systematic study of the oscillatory  nanofluidics of single-crystal diamond nanomechanical resonators. We  explore a broad parameter space, focusing on both the resonator length scale (size)  and the resonance frequency. Previous works typically focused on only one of these parameters, i.e., either the frequency \cite{Karabacak, Universality,Ekinci_LOC,Yakhot_Colosqui} or the length scale \cite{Bullard}. We show conclusively how  a subtle  interplay between the length scale and the frequency determines the nature of the flow induced by nanomechanical resonators   --- resulting in  low-frequency and high-frequency regimes. We also  compare the surface accommodation coefficient of heavy (N$_2$, Ar) and light (He) gases on diamond resonators and determine that diamond surface accommodates these gases differently. Finally, we  measure the thermal fluctuations of the nanomechanical resonators in water and compare these measurements with theory \cite{Paul_PRL, Paul_Nanotechnology, Paul_JAP}.

\section{Experiments}

Figure 1a shows scanning electron microscopy (SEM) images of a set of diamond nanomechanical resonators. To make these devices,  we use a method similar to the angled-etching fabrication technique described in \cite{Burek_2}. This technique  was developed due to a lack of a mature thin-film technology for depositing single-crystal diamond. Briefly, we first perform a standard vertical etch using oxygen plasma, with a  second etch step done at an oblique angle.  Figure 1b shows the sidewalls of one of the cantilevers; the inset is a cross-sectional image. To take these images, the diamond nanocantilevers are transferred onto an evaporated silver film by flipping the chip, pressing the chip on the film, and manually breaking the nanocantilevers. The sidewall images are taken  after these steps. For the cross-sectional images, the sample is further coated with  a few-micron-thick  platinum layer and cut through by a Focused Ion Beam (FIB) tool. The  image  is  taken at a tilt angle of 52$^o$ and is subsequently tilt-corrected.

Returning to Fig. 1a, we identify several interesting features. Due to  angled-etching, the cross-sections of the cantilevers are not rectangular but  triangular. Figure 1b shows that most of the sidewall surfaces are  smooth, with an estimated root-mean square (r.m.s) roughness $\lesssim 10$ nm. The top surface of the cantilever is protected during etches and is much smoother, with an r.m.s. roughness $<1$ nm. In Table 1, the approximate linear dimensions ($l \times w \times h$) of the nanocantilevers  measured from SEM are listed. In addition, there is a  6-$\mu\rm m$ gap between the nanocantilevers and the substrate.  Crucial to our study are the length $l$ and the width $w$. The length $l$ primarily determines the vacuum resonance frequency $f_0={\omega_0 \over 2\pi}$ here, since $w$ and $h$ are the same for all our devices. The width  $w$ is the relevant length scale for the flow:  for the cross flow generated by the out-of-plane (along the $z$-direction) oscillations of the cantilever, the cantilever can be approximated as a cylinder of diameter $w$.

\begin{table*}
\caption{\label{tab:devices} Linear dimensions, vacuum parameters, transition pressure $p_c$, and water parameters. Missing entries correspond to devices that could not be measured in water. The spring constant is determined as $k_0 \approx \frac{{Ew{h^3}}}{{{12l^3}}}$ \cite{Cleland}. For the cantilevers for which the thermal amplitudes were measured,  calculated $k_0$ values agreed  (to within $\sim 50 \%$) with the experimental values determined from the equipartition of energy \cite{Cleland}.}
\begin{ruledtabular}
\begin{tabular}{ccccccc}
$l \times w \times h$ & $f_0$& $k_0$ & $Q_0$ & $p_c$& $f_w$ & $Q_w$\\
 ($\mu \rm m^3$) & (MHz) & (N/m) &  & (Torr) & (MHz) &  \\
\hline
$48\times0.820\times0.530$ & $0.411$ &$0.11$ & $1.4\times10^5$ & $44\pm3$ &$-$&$-$\\
$43\times0.820\times0.530$ & $0.539$ &$0.15$ & $1.5\times10^5$ & $46\pm3$ &$-$&$-$\\
$38\times0.820\times0.530$ & $0.686$ &$0.22$ & $1.45\times10^5$& $47\pm3$ &$0.21$&0.75\\
$34\times0.820\times0.530$ & $0.894$ &$0.31$ & $1.2\times10^5$ & $56\pm4$ &$0.29$&0.95\\
$29\times0.820\times0.530$ & $1.211$ &$0.53$ & $1\times10^5$& $44\pm6$ &$0.43$&1.1\\
$24\times0.820\times0.530$ & $1.735$ &$0.89$ & $1\times10^5$& $50\pm4$ &$0.73$&1.3\\
$19\times0.820\times0.530$ & $2.691$ &$1.8$ & $1\times10^5$   & $50\pm4$ &$1.30$&1.55\\
$14.7\times0.820\times0.530$ & $4.725$ &$4$ & $9.2\times10^4$ & $52\pm4$& $2.55$&2\\
$9.6\times0.820\times0.530$ & $10.421$ &$14$ & $7.7\times10^4$ & $64\pm5$&$-$&$-$\\
$9.6\times1.300\times1.150$ & $28.653$ &$228$ & $2.1\times10^4$ & $187\pm15$&$-$&$-$\\
$4.8\times0.820\times0.530$ & $40.032$ &$112$ & $4.7\times10^4$ & $211\pm11$&$-$&$-$\\

\end{tabular}
\end{ruledtabular}
\end{table*}

The measurements of the out-of-plane (along $z$ in Fig. 1a) displacements of the nanocantilevers are performed using a path-stabilized Michelson interferometer. The displacement sensitivity of the interferometer is estimated to be $\sim 30~\rm fm /\sqrt Hz$ in the range 1-50 MHz with {40 $\mu \rm W$} incident on the photodetector. The displacement sensitivity becomes worse at low frequencies due to technical noise (see Fig. 4a below). The optical spot  is $\approx 1 ~\mu \rm m$ in diameter, and the typical power incident on a nanocantilever is {$\sim100 ~\mu \rm W$}.  For measurements in gases, we use a home-built vacuum chamber which can attain a base pressure of $p\approx 10^{-7}$ Torr after being pumped down by an ion pump. The chamber is fitted with calibrated capacitive gauges for accurate and gas-independent pressure measurements. For the experiments in water, we use a  small fluid chamber, filled with water and sealed with a cover slip \cite{Sukumar_PRL}.

In our experiments with gases, we monitor the dissipation of the nanomechanical resonators as a function of the  surrounding gas pressure $p$. The gases used in these experiments are  high-purity He, N$_{2}$, and Ar. The (dimensionless) mechanical dissipation $\frac {1} {Q_m}$ is obtained by fitting the thermal fluctuations  or the driven response of the nanomechanical resonators. To find the fluidic dissipation $\frac {1} {Q_f}$, we subtract the intrinsic dissipation $\frac {1} {Q_0}$ obtained at the base pressure from the measured dissipation: $\frac {1} {Q_f} = \frac {1} {Q_m} - \frac {1} {Q_0}$.  Figure 1b shows the power spectral density (PSD) $S_z(f)$ of the thermal fluctuations of a nanocantilever with dimensions $29 \times 0.820 \times 0.530~\mu \rm m^3$  immersed in  N$_{2}$ at atmospheric pressure. Note that the dissipation increases but the resonance frequency does not shift significantly when going from vacuum to atmosphere (see Table 1). Figure 1c shows $S_z(f)$ for the same device in water. The changes observed in going from atmosphere to water are quite dramatic.

\section{Results and Discussion}
\subsection{Gas: Dissipation Scaling and Dimensionless Numbers}

\begin{figure*}
\centering
\includegraphics[width=6.75in]{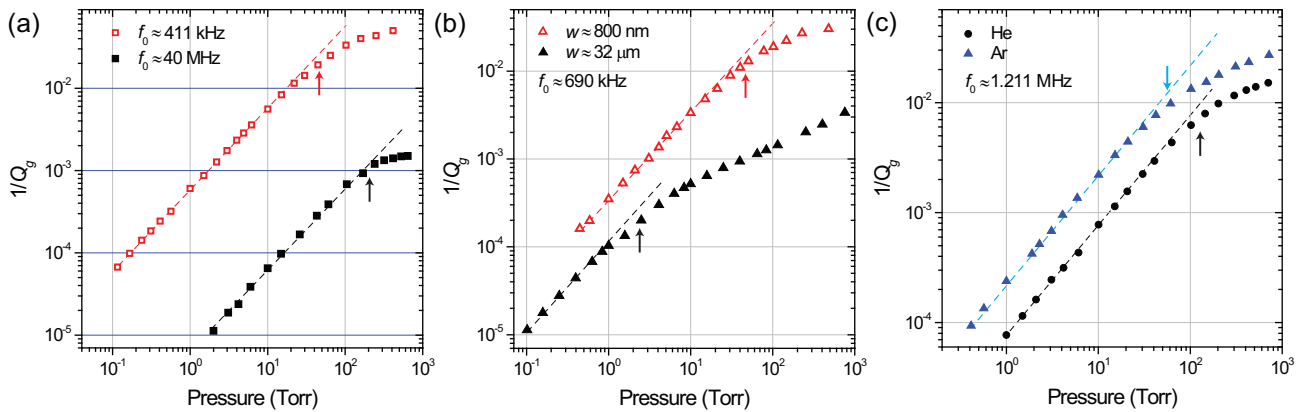}
\caption{Gas dissipation $1/Q_g$ \emph{vs.} pressure $p$ while exploring different parameters. Dashed lines show the molecular flow regions with $1/Q_g \propto p$. The arrows indicate the critical pressures $p_c$ around which the flow transitions from the molecular flow regime into the viscous regime. (a) Fixed linear dimension $w \approx 820$ nm but two different frequencies, $f_0 \approx 40$ MHz and $f_0 \approx 411$ kHz. (b) Fixed $f_0 \approx 690$ kHz but different linear dimensions. The lower data trace is from a rectangular silicon microcantilever ($130 \times 32.5 \times 1~\mu {\rm m}^3$) while the upper one is from a  diamond nanocantilever ($ 38 \times 0.820 \times 0.530~\mu {\rm m}^3$).  (c) Same cantilever ($f_0 \approx 1.211 $ MHz) but two different gases, He and Ar. The error bars in all data points are smaller than symbol sizes.}
\label{fig:Qgas}
\end{figure*}

We first establish the basic aspects of cantilever-gas interactions. At very low pressures, the mean free path $\lambda$ in the gas is large. The problem of a cantilever oscillating  in the gas can be simplified by assuming that gas molecules  collide only with the cantilever surfaces  but not with each other.  At this limit, the  dissipation can be  found to be ${1\over Q_g} \propto p$ (see Eq.~({\ref{Christian_formula}}) below) \cite{Christian, Gombosi}. At the opposite limit of high pressures, the dissipation eventually converges to another asymptote, ${1 \over Q_g} \propto p^{1/2}$. This is the continuum limit in which  the Navier-Stokes equations are to be used \cite{Sader}. The crossover between these two asymptotes (the transitional flow regime)  manifests itself as a gradual slope change in the dissipation \emph{vs.} pressure data. The pressure $p_c$, around which this transition occurs, is therefore a fundamentally important parameter and should provide  insights on the scaling of this flow problem.   Below, we investigate the transition pressures  $p_c$ for different  nanocantilevers and extract the  physical dimensionless parameters relevant to the problem. Because the gaps  are large in our samples,  squeeze damping becomes mostly irrelevant here \cite{Lissandrello, Bhiladvala_Knudsen, Bao_Squeeze}. We also do not observe any deviations from the linear $p$ dependence in the  molecular flow (low-pressure) region, in contrast to the observation  at low temperature (4.2 K) reported in \cite{Defoort}.

Examples of our gas dissipation  $1 \over Q_g$ measurements as a function of pressure $p$  are shown in Fig. 2, with the dashed line being the asymptote proportional to $p$. We  determine the transition pressure $p_c$ consistently in all experiments by finding the pressure at which the dissipation deviates by  $25\%$ from the low-$p$ asymptote.  Figure 2a shows    $1 \over Q_g$ in N$_2$ for two different cantilevers, which possess identical cross-sections ($w\times h$), but different lengths and resonant frequencies. The  linear dimension relevant to the flow here is the  width  $w$ \cite{Sader}, and it is kept the same. Yet, the cantilevers with $f_0 \approx 40$ MHz and $f_0 \approx 411$ kHz go through transition at pressures $p_c \approx 211 $ Torr and   $p_c \approx 44$ Torr, respectively. Thus, the  frequency of the flow appears to be the relevant physical parameter for this case. In Fig. 2b, we explore the opposite limit by comparing one of our diamond devices with a commercial silicon microcantilever with linear dimensions $130 \times 32.5 \times 1~\mu \rm m^3$. Both the nanocantilever and the microcantilever have the same  frequency, but the width of the microcantilever is much larger. In this limit, the microcantilever  attains transitional flow at a significantly lower pressure ($p_c\approx 1.5$ Torr) as compared to the nanocantilever ($p_c\approx 47$ Torr).  In Fig. 2c, we further investigate the relevance of the length scale by measuring the same device in Ar and He.  The mean free path  $\lambda$ depends upon the diameter of the gas molecules, $\lambda \approx 0.23\frac{k_{B}T}{d^{2}p}$; and $d_{\rm He} \approx 2.20$ {\AA} and $d_{\rm Ar} \approx 3.64$ {\AA} \cite{Hanlon}. Although subtle, it can be seen that the $p_c$ value in  Ar is less than that in He. We also note that, at a given pressure, the dissipation in Ar is larger than that in He by a factor $\approx \sqrt {m_{\rm Ar}/m_{\rm He}}$,  where $m$ is the molecular mass (see discussion below and Eq.~(1)).

The data in Fig. 2 allow for  a dimensional analysis of the problem,  with the two relevant parameters being the resonance frequency and the length scale. Recently in a series of experimental \cite{Karabacak, Universality}, theoretical \cite{Yakhot_Colosqui}, and numerical \cite{Colosqui_LB} studies, we have shown that one needs to  use  the dimensionless frequency (or Weissenberg number),  ${\rm Wi} = 2 \pi f_0 \tau=\omega_0 \tau$, in order to characterize a flow oscillating at frequency $f_0$ in a fluid with relaxation time $\tau$. In particular, we and others \cite{Svitelskiy,TJ_cantilever} have observed that the characteristic length scale of the flow (resonator) drops out of the problem, provided that the flow is in the ``high-frequency limit". Then, the transition from molecular flow ($\omega_0 \tau \gg 1$) to viscous flow ($\omega_0 \tau \ll 1$)  takes place  when $\omega_0 \tau \approx 1$. For  a near-ideal gas, $\tau\propto {1\over p}$ \cite{reif}, indicating that $p_c$ should scale with frequency as ${{f_0} \over {p_c}} =\rm {constant}$. Indeed, the gray data points (gray squares) in Fig. 3a from earlier work \cite{Karabacak} on \emph{micro-} and \emph{nanomechanical} resonators show a linear relation between $p_c$ \emph{vs.} $f_0$.  We plot the $p_c$ \emph{vs.} $f_0$ values for the diamond nanocantilevers (large open diamonds) on top of our earlier data \cite{Karabacak}  in Fig. 3a. Surprisingly, the linear trend between $p_c$ and $f_0$ holds only for high frequencies, with a saturation at low frequencies. In other words, resonance frequency  is the relevant parameter that determines the transition, but only above a certain frequency.   When the frequency is low enough, we deduce that the  length scale and hence the Knudsen number ${\rm Kn} = \lambda / w $  must emerge as the dominant dimensionless number. Indeed, the data appear to saturate when $\lambda\sim w \sim 800$ nm, indicating that the transition from molecular flow ($\rm Kn \gg 1$) to viscous flow ($\rm Kn \ll 1$) takes place around ${\rm Kn} = \lambda / w \approx 1$. This is a novel observation. Returning to our earlier (gray) data,  we explain why this deviation is not present there \cite{Karabacak}: there, the high-frequency data were obtained on nanomechanical beams with $w \sim 500$ nm (pink region in Fig. 3a), and the low-frequency data  were obtained on microcantilevers with $w \approx 30~ \rm \mu  m$ (blue region in Fig. 3a). Because ${\rm Kn} = \lambda / w $ was small for all the resonators (see $\lambda$ values on the right $y$-axis), the flows remained in the high-frequency limit in these previous experiments.

We conclude that the physics of the flow around a mechanical resonator  must be determined by an interplay between the  size and the frequency of the resonator. In order to show this more quantitatively, we scrutinize    ${\rm Kn} = \lambda /w$ and  ${\rm Wi} = \omega_0 \tau$ for each  device at its transition pressure $p_c$, taking into account the differences between gases. The relaxation time for N$_2$ as a function of $p$, $\tau_{\rm N_2} \approx {\rm constant}/p$, is available empirically  from \cite{Karabacak}, i.e., the line in Fig. 3a. We determine the  relaxation times of the He and Ar using kinetic theory, e.g., ${{{\tau _{{\rm{Ar}}}}} \over {{\tau _{{\rm{N_2}}}}}} \approx {\left( {{{{d_{{\rm{N_2}}}}{\rm{ }}} \over {{d_{{\rm{Ar}}}}}}} \right)^2}\sqrt {{{{m_{\rm N_2}}} \over {{m_{\rm Ar}}}}}$. We  plot  ${\rm Kn} = \lambda /w$ and  ${\rm Wi} = \omega_0 \tau$ in the $xy$-plane in Fig. 3b. If  $\rm Kn$ is small, $\rm Wi \approx 1$  and becomes the dominant parameter --- and \emph{vice versa}. This suggests that the dissipation should be a function of both the frequency and the length scale: $Q_g = Q_g ( {\rm Wi} , {\rm Kn})$. While more theoretical work is needed to obtain the function, the results in Fig. 2 and 3 capture the essence of the physics.


\begin{figure}

\includegraphics[width=3.375 in]{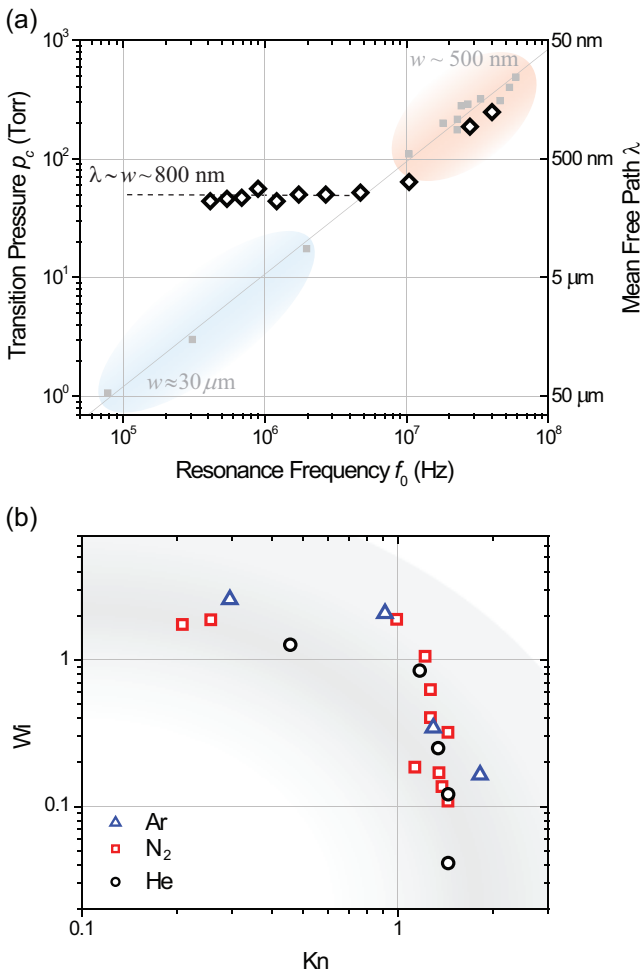}
\caption{(a) Transition pressure $p_c$ as a function of resonance frequency $f_0$ for the diamond cantilevers in this work (large open diamonds). For comparison, earlier data from \cite{Karabacak} are also shown  (gray squares). The gray line is ${{p_c}} = {\rm {constant}} \times {f_0}$ and corresponds to the linear fit in \cite{Karabacak}, $\tau_{\rm N_2} = {{1.85 \times 10^{-6}~{\rm [s-Torr]}} \over p}$. The blue and pink  regions distinguish between the data obtained on microcantilevers (blue) from data on nanomechanical beams.  The approximate characteristic dimensions $w$ of these resonators are as indicated. (b) ${\rm Wi} =\omega_0 \tau$ \emph{vs.} ${\rm Kn} =\lambda /w$ for the diamond nanocantilevers at the transition pressure $p_c$ in He, N$_2$, and Ar.}
\label{fig:tauP2}
\end{figure}

\subsection{Gas: Surface Accommodation}

We next focus on the microscopic interaction between the gas molecules and the single-crystal diamond surface. The  molecular flow (low-pressure) regime provides interesting clues on how gas molecules bounce back from  a solid surface.  Using kinetic theory, one can derive an approximate formula  for  the dissipation  \cite{Martin, Gombosi, Bullard} during the out-of-plane oscillations of a thin plate:
\begin{equation}
{1 \over Q_g} \approx { m^{1/2} p \over 2 \pi K(\epsilon) \rho_s h  f_0  (2k_B T)^{1/2}}
\label{Christian_formula}
\end{equation}
Here, $m$ is the  mass of a gas molecule; $\rho_s$ is the density of the plate; $h$ and $f_0$ are the plate thickness and oscillation frequency, respectively; $k_B$ is the Boltzmann constant and $T$ is the temperature. The function $K(\epsilon)= {{\sqrt \pi} \over {4 + \pi  + (4 - \pi )\epsilon }}$ depends upon the parameter $\epsilon$, where $\epsilon = 0$ for diffuse reflections and $\epsilon =1$ for specular reflections \cite{Bullard}. $K(\epsilon)$ changes by about $10~\%$, i.e., ${K(1) \over K(0)} \approx {0.22 \over 0.25} \approx  0.88$, if all the reflections are specular instead of diffusive.  Because some reflections are specular and others are diffusive,  an average $K$ value emerges for each surface \cite{Ducker_2010}. In principle, Eq. (1) allows for determining   $K$ for a given cantilever  --- provided that all the quantities in the equation can be measured with high accuracy.


\begin{figure*}
\centering
\includegraphics[width=6.75in]{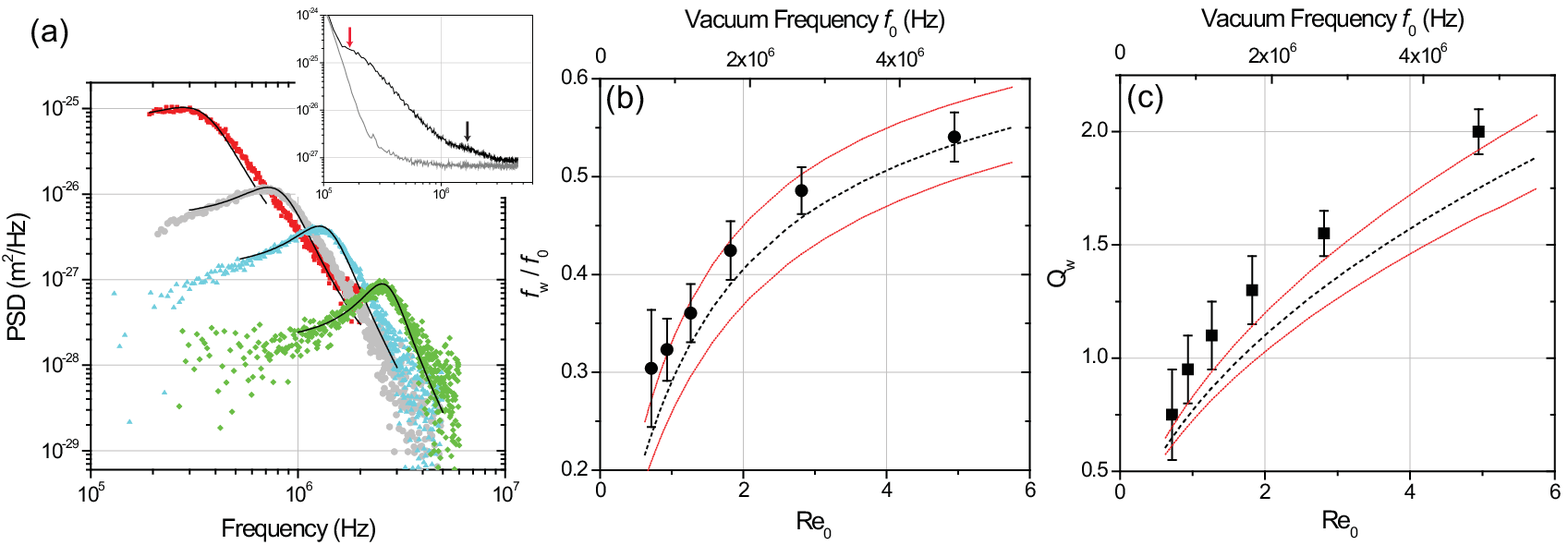}
\caption{(a) Power spectral density (PSD) of the thermal fluctuations of four nanocantilevers with $f_0 \approx 0.894, 1.735, 2.691$ and 4.725 MHz in water. The solid lines are fits to Eq.~(3). The inset shows  raw data for a cantilever with $f_0 \approx 539$ kHz. The black data trace is measured at the tip of the cantilever and the gray at the base. The fundamental-mode peak (red arrow) cannot be fully resolved; the slight peak (black arrow) around 2 MHz is the first harmonic mode of the cantilever. Technical noise dominates the measurement for $f \lesssim 10^5$ Hz; the measurement is shot noise limited at high frequencies. (b) The ratio of the peak frequency in water to that in vacuum, $f_w/f_0$,  as a function of  Reynolds number ${\rm Re_0}$. The upper $x$-axis is $f_0$. (c) Quality factor $Q_w$   in water. The (black) dashed lines are predictions of \cite{Paul_Nanotechnology}. The dotted lines indicate the upper and lower bounds of theory based on the uncertainty in cross-sectional dimensions. }
\label{fig:water}
\end{figure*}

Here, we will not attempt to determine the absolute value of $K$ for each gas. Instead, however, we can obtain  $K$ ratios in different gases assuming Eq.~(1) applies:
\begin{equation}
K_{\rm Ar}:K_{\rm N_2}:K_{\rm He} \approx 1:0.97\pm 0.01:0.92\pm 0.01.
\end{equation}
This result is obtained as follows. For a given  resonator, the low-$p$ region of the dissipation data in Ar, N$_2$ and He are fit to lines, providing three different slopes for Ar, N$_2$ and He (see Fig. 2c). The slopes from this resonator are then used to form the ratio in Eq. (2). Because all the factors in Eq. (1) including $h$ are divided out, this operation isolates the effect of the gas. The experiment is then repeated for other resonators, and the $K$ values from different resonators are averaged. The data in Eq. (2) above come from five different resonators.

Our results in Eq.~(2) suggest that He reflects more specularly than heavier gases, Ar and N$_2$; and that Ar and N$_2$ behave similar to each other. These facts are  not unexpected and  qualitatively agree  with earlier observations on other crystalline surfaces \cite{Micro_Nanoflows_Book}. What is perhaps surprising is that most He particles appear to reflect specularly from the diamond surface. In fact, any roughness on the surface will tend to reduce the fraction of particles that are reflecting specularly \cite{Seo_Ducker_PRL,Seo_2014,Sedmik_2013}, and our surfaces have r.m.s. roughness $\lesssim 10$ nm, as noted above.  In order to modify Eq.~(1) to accurately incorporate the effect of roughness in $K$, the spatial  distribution (dominant wavelengths) of roughness ought to be taken into account,  in addition to its r.m.s. value. Experimentally, it is a current technical challenge to measure the distribution of roughness on nanostructures such as ours.

\subsection{Water}

Finally, we turn to the fluid dynamics of nanomechanical cantilevers in water. Figure 4a shows the power spectral densities (PSDs) of the thermal fluctuations of four  nanocantilevers with   $f_0 \approx 0.894, 1.735, 2.691$ and 4.725 MHz in water. The  PSDs in water can be improved by a noise subtraction process. We  illustrate this process by turning to the raw data for a low-frequency cantilever ($f_0 \approx 539$ kHz) shown in the inset of Fig. 4a: the lower (gray) trace is the PSD recorded with the optical spot at the base of the cantilever; the black trace is taken at the tip of the same cantilever. Because noise powers are additive and the interferometer gain remains the same between the two noise traces, one can subtract the noise power at the base from the noise power at the tip at each frequency to obtain the PSD of the mechanical fluctuations. The two arrows in the inset show  the peaks of the fundamental mode and the first harmonic mode of the nanocantilever. Technical noise at low frequencies (e.g., laser noise) obscures the fundamental-mode peak, making it impossible to obtain a fit (hence, the missing entries at low frequencies in Table 1).

In order to extract the device parameters, we  fit the peaks in the PSDs to Lorentzians (solid lines in Fig. 4a):
\begin{equation}
{S_z}(\omega ) \approx {{4{k_B}T{\gamma _w}} \over {{m_w}^2}} \times {1 \over {{{\left( {{{{k_0}} \over {{m_w}}} - {\omega ^2}} \right)}^2} + {\omega ^2}{{{\gamma _w}^2} \over {{m_w}^2}}}}.
\end{equation}
Here, $m_w$ is the mass of the entrained water added to the mass of the cantilever; $k_0$ is the spring constant of the cantilever; and $\gamma_w$ is the dissipation. We treat  $k_0 \over m_w$ and $\gamma_w \over m_w$ in Eq.~(3) as frequency-independent fitting parameters and adjust the amplitude (${{4{k_B}T{\gamma _w}} \over {{m_w}^2}}$) freely for acceptable fits. (In reality,  $\gamma_w$ and $m_w$ are both frequency dependent \cite{Paul_PRL}.) This exercise provides the quality factor $Q_w$ and the peak frequency $f_w$ in water, where  $Q_w \approx2 \pi f_w  {m_w \over \gamma_w}$ and  $ f_w \approx 2\pi \sqrt{k_0 \over m_w} \sqrt {1-{1 \over 4{Q_w}^2}}$. The highest frequency cantilever in Fig. 4a can be fit accurately; however, the fits progressively become  worse with decreasing $f_0$. In particular, the low-frequency tail in each data set is hard to reproduce in the fit  and contributes to the  reported errors. Regardless, we extracted  $f_w$ and $Q_w$ values  for all the data sets in which a fundamental peak could be resolved (for instance, the peak cannot be fully resolved in the inset of Fig. 4a). The extracted frequency ratios $f_w/f_0$ and quality factors $Q_w$ are plotted in Fig. 4b and 4c, respectively, as a function of the dimensionless frequency parameter (or the frequency-dependent Reynolds number) defined as ${\rm Re_0} = { 2\pi f_0 w^2 \over 4 \nu}$ \cite{Paul_PRL,Sader}, with $\nu$ being the kinematic viscosity of water. The upper $x$-axes in Fig. 4b and 4c  display the vacuum frequencies $f_0$.  The error bars correspond to the  parameter range that provided acceptable fits. As $f_0$ is reduced, the quality factor  $Q_w$ in water goes down;  the cantilever response eventually becomes overdamped ($Q_w <1/2$).

The dashed lines in Fig. 4b and 4c are the predictions of a theory developed by Paul and co-workers \cite{Paul_PRL, Paul_Nanotechnology, Paul_JAP} for describing the thermal fluctuations of a nanocantilever in a fluid. The  uncertainty in the cross-sectional dimensions of the nanocantilevers result in  the two (red) dotted lines which indicate the upper and lower bounds of the predictions.  In the approach of Paul and co-workers, one first determines the frequency-dependent fluidic dissipation by approximating the nanocantilever (or nanobeam) as an oscillating cylinder; one then uses the fluctuation-dissipation theorem to obtain the frequency spectrum of the nanocantilever fluctuations from the dissipation.  To compare experiment and theory, one needs the mass loading parameter $T_0 $ \cite{Paul_PRL, Paul_Nanotechnology, Paul_JAP}, which is the ratio of the mass of a cylinder of fluid of diameter $w$ to the actual mass of the solid. We compute the $T_0$ value from cross-sectional images of the nanocantilevers, such as the one shown in Fig. 1b. Considering the uncertainty in the cross-sectional dimensions and the non-uniformity of the cross-section along the length of the nanocantilever, we obtain the average value $T_0= 0.69 \pm 0.1$.    The theory accurately predicts  both the quality factors and the peak frequencies in water for this $T_0$ range. The slight disagreements could be the result of the cylinder approximation: the cantilevers are triangular in cross-section; furthermore, the 2-D flow  (long cylinder) approximation  assumed in the theory may  be introducing errors. The presence of the substrate may be another complicating factor.  Extending the frequency range of the experiments, especially towards lower frequencies, may result in a better understanding of the limits of the theory. Recent studies \cite{Jeney_Nature} on optically-trapped micro-particles, for instance, have uncovered novel effects  due to the coherent   hydrodynamic memory of the liquid during thermal oscillations.

\section{Conclusion}

In conclusion, we have thoroughly investigated the fluid dynamics of single-crystal diamond nanocantilevers. Several aspects of our results are noteworthy. First, we show that a competition between two dimensionless numbers, one associated with the cantilever size ($\rm Kn$) and the other with the resonance frequency ($\rm Wi$),  determines the dissipation in a gas. While our dimensional analysis explains the physics, more theoretical development is needed for obtaining an analytical expression for the dissipation. Second, to the best of our knowledge,  this is the first report of the measurement of the accommodation coefficient on single crystal diamond; our measurement approach using nanomechanical resonators is also novel. This  approach could be improved to obtain the absolute value of the accommodation coefficient: dissipation  on a properly cooled resonator, on which gas molecules are likely to  stick, and that on a room temperature resonator can be compared --- in a way similar to what we did for different gases. Finally, our studies in water  clarify both the challenges and the  open fundamental questions. From a device perspective,  both the quality factor and the peak frequency in water decrease with decreasing vacuum frequency. The high frequency limit in water, recently observed for breathing modes of nanoparticles \cite{Sader_Visco_1} and nanowires \cite{Sader_Visco_2}, is not yet accessible by the flexural modes of nanofabricated structures like ours.


\begin{acknowledgments}
Device fabrication was performed in part at the Center for Nanoscale Systems (CNS), a member of the National Nanotechnology Infrastructure Network (NNIN), which is supported by the National Science Foundation under NSF award no. ECS-0335765. CNS is part of Harvard University. The Harvard team acknowledges the financial support from STC Center for Integrated Quantum Materials (NSF
grant DMR-1231319) and the Defense Advanced Research Projects Agency (QuASAR program).  The authors gratefully acknowledge Dr. Andrew Magyar for taking cross-sectional images of the device, Dr. Charles Lissandrello for a critical reading of the manuscript, and Prof. Mark Paul for discussions. The authors declare no competing financial interest.
\end{acknowledgments}


\end{document}